\documentclass{rrparticle}
\usepackage{graphicx}

\newcommand{\miktex}{\hbox{Mik\kern-.15em\TeX}}

\title{Isochoric cooling of air in the University Physics laboratory} 
\author[1]{Ivan Z. Stefanov}
\author[2]{Sava Donkov}
\author[3]{Nikolay Denev}

\affil[1]{Department of Applied Physics, Technical university of Sofia, \authorcr 8, Snt. Kliment Ohridski Blvd.,   1000 Sofia, Bulgaria, \authorcr Email:{\em izhivkov@tu-sofia.bg}}
\affil[2]{Institute of Astronomy and NAO, Bulgarian Academy of Sciences, 72 Tsarigradsko Chausee Blvd., 1784 Sofia, Bulgaria, \authorcr Email:{\em sddonkov@astro.bas.bg}}
\affil[3]{Department of Applied Physics, Technical university of Sofia, \authorcr 8, Snt. Kliment Ohridski Blvd.,   1000 Sofia, Bulgaria, \authorcr Email:{ \em ndenevtph@tu-sofia.bg}}
\DeclareUnicodeCharacter{2212}{-}
\begin{document}
	\maketitle
\begin{abstract}
 In this paper an alternative way of conducting the physics laboratory exercise for determining the adiabatic index of air using the Clement -- Desormes method is proposed.
 The process of isochoric cooling of air has been studied in terms of the dependence of pressure on time, and hence temperature on time, since it is proportional to pressure at a constant volume and mass of air. A theoretical model of the considered process was also made. The experimental results were processed statistically. The coefficient of determination $R^2$ and the F-test statistic were calculated and their values indicate a very good agreement between theory and experiment. The analysis of the residuals, however, implies that the model could be further improved through the inclusion of higher order terms.
\end{abstract}
\section{Introduction}
One of the main topics for laboratory exercises from the university course in physics, section "Thermodynamics", is that of determining the adiabatic index of air by the method of Clement - Desormes, \cite{Sivukhin}. In this experiment, two isochoric processes and one adiabatic process take place.
We propose an alternative laboratory exercise using the same experimental setup. We concentrate on one of the thermodynamic processes –- the isochoric cooling of air. We suggest that the reading of the pressure gauge data be done by video recording. The mobile phone camera is used for this purpose. The reasons for this choice are as follows. Every student has a mobile phone. Lately, even the cheaper phones are equipped with a good enough camera, capable of taking good quality pictures. It becomes convenient and fast. It saves time in comparison to the classical measurement method. Moreover, the use of mobile phone cameras for laboratory measurements becomes more and more popular among students and continues to draw the attention of researchers \cite{MobilePhone, Smartphone2020, SmartphoneAmJPh, Smartphone2022}. Usually, when they need to record some information, especially  in the case of large data sets, students immediately take a picture with their phone. Another positive side of the proposed method is that it is convenient for distance learning. It should also be emphasized that this is an alternative method that can increase the interest of students.
The presented laboratory exercise involves a developing of a nontrivial theoretical model which including the first principle of thermodynamics, the definition of heat capacity, the equation of state of an ideal gas, the heat transfer equation. In addition, here we also have air leaks from the system that might affect the results and should be commented on by the instructor.

After the recording, the data is read off the video clip and filled in an Excel spreadsheet. The use of spreadsheets data analysis, \cite{SpreadsheetPhysics, Excel} is increasingly popular. We do the same. After statistical processing, the results are compared with the theoretical model.

The article contains an introduction, a brief description of the classical method, a description of the alternative method, a theoretical model, a statistical treatment of the results and a comment on their coincidence with the theoretical model, a conclusion.

\section{Experiment}
\subsection{Traditional approach}
\begin{figure}[ht!]
	\centering
	\includegraphics[width=0.9\textwidth,keepaspectratio]{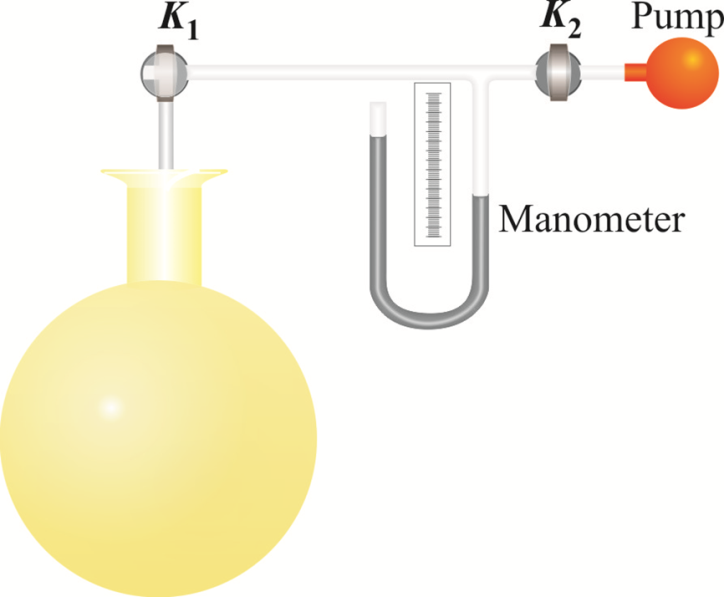}
	\caption{The experimental setup. \textit{(Author of the drawing: assoc. prof. Mihail Mihalev)}}\label{Setup}
\end{figure}

In the classical Clement-Desormes method, the following algorithm is used.  Initially valve $K_1$ is closed, while $K_2$ is open. Then, with the help of a rubber pump, additional air is introduced into the working volume, Fig.~\ref{Setup}, and so the pressure there increases relative to the atmospheric pressure. The temperature also rises compared to room temperature. After that, tap $K_2$ is immediately closed and the air in the working volume begins to cool isochorically until it reaches room temperature. During this time, the change in pressure in the system relative to atmospheric pressure is monitored using a water manometer, one leg of which is connected to the working volume and the other is open to the atmosphere. When the level difference in the two legs of the manometer stops changing, then we consider that the temperature in the system has equalized with the room temperature and the pressure in the working volume no longer changes. This level difference is written as $\Delta h_1$.
Then we perform an adiabatic process by quickly opening and closing tap $K_1$. In this process we have a rapid expansion of the air until it reaches atmospheric pressure and a temperature lower than room temperature. This is followed by a process of isochoric heating of the air in the working volume until it reaches room temperature. Again, the difference in levels in the two legs of the manometer is monitored and when it stops changing, we record it as $\Delta h_2$.
The adiabatic index for air is determined from the values of $\Delta h_1$ and $\Delta h_2$ using a simple formula.
The entire procedure is repeated 10 times and the average value of the adiabatic index and measurement error are reported. This takes a considerable amount of time, as one has to wait for equilibrium to be reached and the relevant isochoric processes to cease.
\subsection{The isochric cooling}
\begin{figure}[ht!]
	\centering
	\includegraphics[width=0.5\textwidth,keepaspectratio]{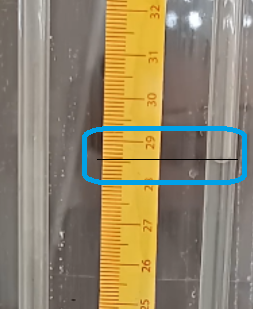}
	\caption{Photo of waer level in on of the legs of the manometer. Its value is read off the scale.}\label{Scale}
\end{figure}

Here our proposal is to study the variation of the air pressure in the working volume with respect to the atmospheric pressure. The procedure is as follows.
With valve $K_1$ closed and $K_2$ open, we pump air into the working volume until a gauge pressure of about 20 cm of water column is reached relative to the atmospheric pressure. After that, we quickly close the valve $K_2$ and start measuring the air pressure in the working volume relative to the atmospheric pressure in an interval of 5 s. The total measurement time is about 3 minutes. Finally we have a table with about 35-36 values for time and pressure.
In order to quickly and accurately measure the values of the water column levels in both legs of the manometer every 5 seconds, we suggest that the measurement be carried out by video recording. It is enough to use the camera of a mobile phone and within 3 minutes to take a picture of the manometer in video recording mode. Then from the clip, by stopping the frame every 5 seconds and zooming in on the image, the values of the water column levels in the two legs of the manometer are taken. See Fig. \ref{Scale}. In order to read the values with good accuracy from the video clip, it is necessary for it to be captured it with good quality and appropriate scale.

\section{Model}

Let us consider a simple theoretical model of the process of isochoric cooling in the working volume. We are dealing with an ideal diatomic gas (air) in the working volume, which can exchange heat with the surrounding medium through the thin glass walls of this volume. The process of isochoric cooling we regard as a quasi-equilibrium thermodynamic process, so the thermodynamic states, during the process obey the equation of Clapeyron-Mendeleev:
\begin{equation}
    \label{Clapeyron-Mendeleev equ}
    PV=nRT~,
\end{equation}
where $P$ is the gas pressure, $V$ is the value of the volume, which is constant, $n$ is the number of molls of the gas in the working volume, $T$ is the Kelvin temperature, and $R$ is the universal gas constant. Since $n= {\rm const}$ during the process in air leakage is neglected, then $P(t)$ and $T(t)$ obey one and the same law as functions of the time $t$. So we intend to study how the temperature varies with time, and finally we can obtain how the $P(t)$ law looks like.

The heat transfer during the isochoric cooling takes place  according to the equation:
\begin{equation}
    \label{heat transfer equ}
    \frac{\delta Q}{dt} = - \kappa S \Delta T~,
\end{equation}
where the meaning of the above quantities is as follows: $\delta Q$ is the infinitesimal heat which the gas in the volume looses during the time $dt$, due to the temperature difference $\Delta T = T(t) - T_0$ (here $T(t)$ is the temperature in the volume and $T_0$ is the temperature of the surrounding medium); $S$ is the surface of the walls, which is obviously fixed; $\kappa= k/d$, where $k$ is the material's conductivity and $d$ is the thickness of the glass walls (both are regarded as constants).

On the other hand for the isochoric process is valid the first law of thermodynamics: $\delta Q = dU + \delta W$, where $dU = n C_{\rm V} dT$ ($C_{\rm V} = 5R/2$ is the thermal capacity of diatomic gas in regard to constant volume) is the infinitesimal increment of internal energy of the gas and $\delta W$ is the infinitesimal work which is done by the gas, both due to the received heat $\delta Q$. Since the process is isochoric, then $\delta W = 0$, and hence $\delta Q = n C_{\rm V} dT$. Combining the latter expression with the equation for the heat transfer we obtain the following differential equation for $T(t)$:
\begin{equation}
    n C_{\rm V} \frac{dT(t)}{dt} = - \kappa S (T(t) - T_0)~,
\end{equation}
and after some simple algebra it can be rewritten in the form:
\begin{equation}
    \label{the equ for T(t)}
    \frac{dT(t)}{dt} + p T(t) = q~,
\end{equation}
where $p = \kappa S/n C_{\rm V}$ and $q = \kappa ST_0/n C_{\rm V}$ are constants of the experiment. The above differential equation is linear and has a simple solution \cite{Riley}, which reads:
\begin{equation}
    \label{solution for T(t)}
    T(t) = C \exp(-pt) + \frac{q}{p} = C \exp \bigg(-\frac{\kappa S}{n C_{\rm V}}t \bigg) + T_0~.
\end{equation}
Here the constant $C$ must be determined by the initial conditions: $T(0) = T_{\rm in}$, hence $C=T_{\rm in} - T_0$ ($T_{\rm in}$ is the highest temperature of the gas, after the end of the pumping).

If one wishes to obtain the law $P(t)$ for  the pressure, this easily can be done by multiplying of the equation for $T(t)$ with the coefficient $nR/V$. Finally one reaches:

\begin{equation}
    \label{solution for P(t)}
    P(t) = (P_{\rm in} - P_0) \exp \bigg(-\frac{\kappa S}{n C_{\rm V}}t \bigg) + P_0~,
\end{equation}
where $P_{\rm in}$ is the highest pressure reached in the volume, and $P_0 = nRT_0/V$ is the pressure of the working gas at the temperature of the surrounding medium and hence it coincides with the atmospheric pressure. So if we are interested in the pressure difference $\Delta P (t) = P(t) - P_0$, then we must expect an exponential law of the form:
\begin{equation}
    \label{equ for delta P}
    \Delta P(t) = (P_{\rm in} - P_0) \exp \bigg(-\frac{\kappa S}{n C_{\rm V}}t \bigg)~.
\end{equation}
In logarithmic scale it represents a straight line with a negative slope.

\section{Reduction of data}
\begin{figure}[ht!]
	\centering
	\includegraphics[width=0.4\textwidth,keepaspectratio]{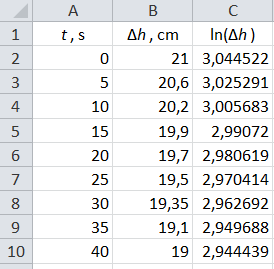}
	\caption{The Excel spreadsheet containing the data.}\label{xlsTable}
\end{figure}

\begin{figure}[ht!]
	\centering
	\includegraphics[width=0.9\textwidth,keepaspectratio]{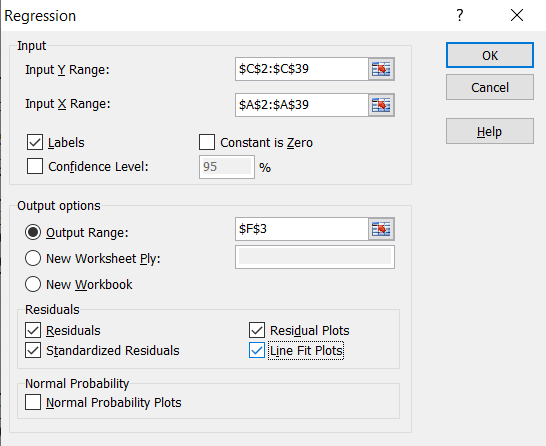}
	\caption{The dialog window in which the options of the ``Regression'' are specified.}\label{Dialog}
\end{figure}

\begin{figure}[ht!]
	\centering
	\includegraphics[width=1.0\textwidth,keepaspectratio]{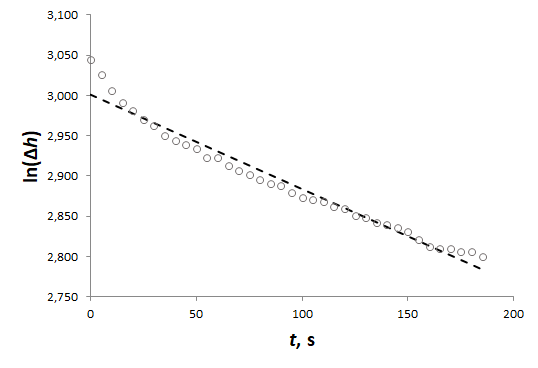}
	\caption{The experimental points and the model line .}\label{experimental data}
\end{figure}

\begin{figure}[ht!]
	\centering
	\includegraphics[width=1.0\textwidth,keepaspectratio]{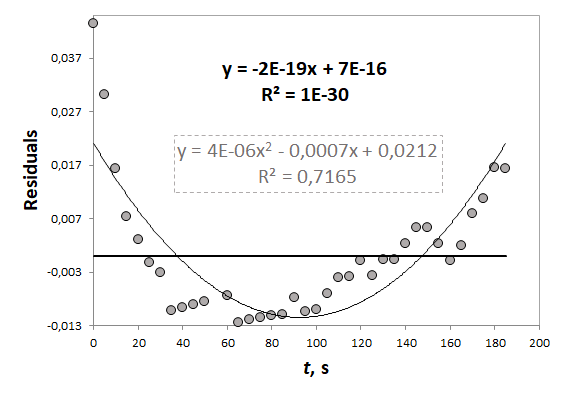}
	\caption{The residuals plus two trend lines -- one of first and and one of second order.}\label{residuals}
\end{figure}

The data can be processed statistically, presented graphically and compared to a theoretical model. We will model the experimental data obtained from the video using a simple regression implemented in Excel. In the first and second columns of the worksheet shown in Fig. \ref{xlsTable}, we fill in the time $t$ and the difference between the water levels in the two legs of the manometer $\Delta h$. As we saw in the previous section, we expect the variation of $\Delta h$ to be described by an exponential law. By taking logarithms, we can obtain a straight line and apply the simple regression method to it. Therefore, the logarithmised values of $\Delta h$ are plotted in the third column.
To apply the simple regression method in Excel\footnote{The regression can be conducted also by other spreadsheet software products.}, we go to the Data menu and select Data analysis, which is located on the right side of the menu. If this function is not visible, it must be enabled from the “Properties” menu. From the dialog box that appears, select “Regression”. A new dialog box appears, shown here in Fig. \ref{Dialog}. 
In it, we must fill in the “Input Y Range” and “Input X Range” by selecting of the corresponding cells from the worksheet. From the “Output options” section, we choose whether the results should be displayed in a new worksheet or in the current one. If we choose the current one, then from the “Output Range” field we must specify the location, i.e. the upper left corner of the table that will contain the results of the statistical processing. We chose the latter. From the “Residuals” section we select all four proposed options: “Residuals”, “Standardized Residuals”, “Residual Plots” and “Line Fit plots”. The last two of these functions allow us to obtain Figs.~\ref{experimental data}~and~\ref{residuals}. To bring them to the form presented here we use the options offered by Excel. For details on obtaining the graphs, we can refer the reader to the following video lecture  \cite{ResidualAnalysisYouTube}. In this same lecture, and also in \cite{SimpleLinearRegressionYouTube, InterpretingLinearRegressionResultsYouTube}, detailed descriptions of the interpretation of the results obtained by the “Regression” function of Excel can also be found. 

On the role of the standardized residuals for the identification of outlier, on the application of the F-test for the estimation of the overall significance of the fit, and in general on simple regression analysis please refer to the following books \cite{StandardizedResiduals, RegressionInExcel, F-test} and video lectures \cite{ResidualAnalysisYouTube, SimpleLinearRegressionYouTube}. 
\begin{figure}[ht!]
	\centering
	\includegraphics[width=1.0\textwidth,keepaspectratio]{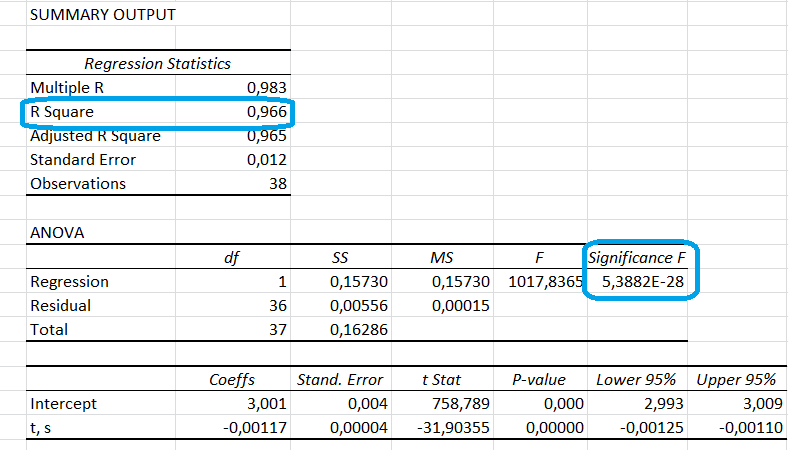}
	\caption{Results of the “Regression” procedure.}\label{Regression results}
\end{figure}
The first criterion we use to evaluate the quality of the approximation, i.e. the extent to which the selected model can explain the obtained experimental data is the $R^2$ coefficient, also known as the "coefficient of determination". Its value is circled in blue in Fig. \ref{Regression results}. Roughly speaking, $R^2=0.966$ means that 96.6\% of the experimental points are well described by the proposed model, which we can say is a pretty good result.
The next estimate of the overall goodness of the fit we can consider is “Significance F”. The value of this parameter is also circled in blue in Fig. \ref{Regression results}. It is practically zero, up to 28 decimal places. This means that the quality of the approximation is very good. In other words, this assessment also does not give us reason to question the applied model.

Figure \ref{experimental data} shows the experimental points and the model line represented by a dashed line. Visually we can establish a good match, i.e. most of the experimental points lie close to the model line, except perhaps the first two or three and the last two.

The residuals from the approximation are shown in Fig. \ref{residuals}. For these, if the model is appropriate, we expect them to be independent and normally distributed around zero. We can find out if this is the case by plotting a trendline as follows. We click with the right button on any of the experimental points from the graph. From the drop-down menu, select "Add Trendline". Then we need to specify whether it is a straight line or a curve of a higher order (up to sixth inclusive). Here we have used a straight line (represented in black color in Fig. \ref{residuals}.) and a parabola (represented in gray color in Fig. \ref{residuals}.). The equations of these curves are also shown in the Figure, each in the corresponding color. What do they tell us? The straight line has zero slope and intercept. These two results testify that indeed the residuals are randomly distributed around the zero value. The coefficient $R^2$ corresponding to the straight line, however, is practically equal to zero, which means that the residuals do not lie near the straight line. The second-order curve describes the residuals significantly better, as evidenced by its coefficient of determination. For the gray line, its value is 0.7165, very close to unity. Therefore, the residuals are not independent, but a systematic curvature is observed, i.e. systematic variation, which is a signal that the model, roughly speaking, can be improved by inclusion of nonlinear terms.

Standardized residuals allow us to detect outliers. These are the points whose standardized residuals are greater than 3. For us, it seems that these are the first two and the last two experimental points.
\section{Conclusion}
This paper proposes a new laboratory exercise for the university course in general physics. The laboratory set-up of the exercise in which the adiabatic index of air is determined according to the Clement-Desormes method, is used. That is, we propose a new way in which one of the common laboratory setups in the standard university general physics laboratory can be used. We focus on one of the processes occurring in the Clement-Desormes experiment, namely the isochoric gas heating process. The exercise proposed here includes an alternative method of measurement using a smartphone, but also modeling, i.e. a theoretical description of the research phenomenon, and finally statistical processing of the experimental results. We establish a good agreement between the proposed model and the obtained experimental results, which is confirmed by two of the three criteria considered: the value of the coefficient of determination, the overall significance of the approximation and properties of the residuals.
We hope that the exercise would be instructive and of interest to students, and that it could be used not only as a laboratory exercise for a standard University Physics course, but can also serve as an idea for a course work or BSc thesis.

\begin{acknowledgement}
The author would like to thank their colleague assoc. prof. Mihail Mihalev, PhD, for drawing the diagram of the experimental setup.
\end{acknowledgement}

\end{document}